\documentstyle[sprocl,psfig]{article}

\def\Journal#1#2#3#4{{#1} {\bf #2}, #3 (#4)}

\def\NPA{{\em Nucl. Phys.} A}
\def\PLB{{\em Phys. Lett.}  B}
\def\PRL{\em Phys. Rev. Lett.}
\def\PRC{{\em Phys. Rev.} C}

\begin{document}

\title{RELATIVISTIC FADDEEV APPROACH TO THE NJL MODEL AT FINITE DENSITY}

\author{M. C. BIRSE, J. A. McGOVERN, S. PEPIN}
\address{Theoretical Physics Group, Department of Physics and Astronomy,
University of Manchester, Manchester M13 9PL, U.K.}
\author{N.R. WALET}
\address{Department of Physics, UMIST, P.O. Box 88, Manchester M60 1QD, U.K.}

\maketitle

\abstracts{
We study the nucleon solution of the relativistic Faddeev equation as a 
function of 
density in the framework of a generalized Nambu--Jona-Lasinio
model. We truncate the interacting two-body channels to the scalar diquark
channel, the coupling constant of which is treated as a parameter. A
3-momentum cut-off is used to regularize the model. The Faddeev equation
is solved numerically using the methods developed by Tjon and others.
At zero density the nucleon is bound only for unrealistically large values 
of the scalar coupling and this binding energy decreases quickly with 
increasing density.}

\section{Introduction}
In contrast to high-temperature zero-density
QCD, rather little is known about high-density zero-temperature
QCD. Due to technical difficulties (the fermionic determinant becoming
complex at finite chemical potential), lattice calculations are not able to
provide unambiguous results. 
However, models of QCD seem to indicate a rich 
phase structure in high density quark matter. In particular, much attention has
recently been devoted to so-called colour superconductivity 
\cite{ARW98,RSSV98}: at high
density, an arbitrarily weak attraction between quarks makes the quark Fermi
sea unstable with respect to diquark formation and induces Cooper pairing of 
the quarks (diquark condensation).
However, the groups who have studied colour superconductivity focused only on 
instabilities of the Fermi sea with respect to diquarks and have not considered
possible 3-quark clustering. To address this question would necessitate in 
principle a generalisation of the BCS treatment. As a first
step we can look for instabilities of the quark Fermi
sea with respect to 3-quark clustering by studying the evolution of the 
nucleon binding energy with density. A bound nucleon at finite density would
be a signal of instability. In this study, we will use the Nambu--Jona-Lasinio
(NJL) model and solve the relativistic Faddeev equation for 
the nucleon as a function of density. 

\section{The model}
The NJL model provides a simple implementation of dynamically broken 
chiral symmetry.
It has been successful in the description of mesonic properties at low energy
and several groups \cite{IBY95,HT94,BAR92} have used it to study baryons at 
zero density (for a review of the NJL model, see for example \cite{Kle92}). 
Several versions of the NJL lagrangian are available. Whatever the version we 
choose, a Fierz transformation allows us to
order the terms according to their symmetries in the $q\bar{q}$ channel; 
here we are only interested in the scalar and pseudoscalar terms:
\begin{equation}
{\cal L}_{\pi} = \frac{1}{2} g_{\pi}[ (\bar{\psi}\psi)^2 - (\bar{\psi}
\gamma_5\tau\psi)^2 ] \quad .
\end{equation}
To study the baryons, another Fierz transformation has to be performed in 
the $qq$ channel. In this work we shall keep only the scalar diquark channel:
\begin{equation}
{\cal L}_s = g_s (\bar{\psi}(\gamma_5 C)\tau_2\beta^A\bar{\psi}^T)(\psi^T(
C^{-1}\gamma_5)\tau_2\beta^A\psi) \quad ,
\end{equation}
where $\beta^A=\sqrt{3/2} \, \lambda^A$ for $A=2,5,7$ projects on the colour 
$\bar{3}$ channel and $C=i\gamma_2\gamma_0$ is the charge conjugation 
matrix. 

The ratio $g_s/g_\pi$ depends on the version of the NJL lagrangian used.
In the following, we will not choose a particular version of the model but 
rather leave the ratio $g_s/g_\pi$ as a free parameter.
We regularize the model with a 3-momentum cut-off $\Lambda$. We 
have two parameters $g_\pi$ and $\Lambda$, which are fitted to the values 
of the pion decay constant $f_\pi=93$ MeV and the constituent quark mass 
$M=400$ MeV. This gives us $g_\pi=7.01$ GeV$^{-2}$ and $\Lambda=0.593$ GeV. 

 The effect of density is introduced by imposing the quark 3-momentum to be 
larger than the Fermi momentum $k_F$. Solving the gap equation as a function
of $k_F$ gives us the usual dependence of the quark constituent mass on 
density. Chiral restoration occurs at $k_F/\Lambda=0.58$, which corresponds 
to about 2.1 times the nuclear matter density 

\section{Diquark at finite density}
 As a first step to the resolution of the Faddeev equation, the scalar diquark 
mass has to be calculated as a function of density. The Bethe-Salpeter 
equation for the 2-body $T$-matrix is solved in the scalar $qq$ channel 
using the ladder approximation (the explicit solution is given 
in \cite{IBY95,HT94}) and the pole of the $T$-matrix gives the 
mass of the bound diquark. Note that the denominator of the scalar $qq$
$T$-matrix is formally identical to that of the pionic $T$-matrix, except
for the replacement of $g_s$ by $g_\pi$. That means that the pion and the 
scalar diquark are degenerate (and of zero mass) for a ratio $g_s/g_\pi=1$.
Fig. 1 gives the binding energy of the scalar
diquark, $B_{diq}=2\sqrt{k_F^2+M^2}- E_{diq}$ as a function of the 
dimensionless variable $k_F/\Lambda$ for a value of the scalar coupling 
$g_s/g_\pi=0.83$.
\begin{figure}
\centerline{\psfig{figure=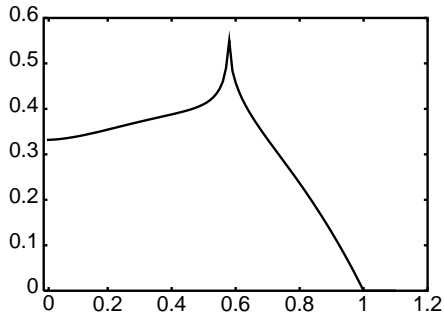,height=4.6cm}}
\caption{Diquark binding energy (in GeV) as a function of $k_F/\Lambda$}
\label{diq}
\end{figure}
One can see that the scalar diquark is bound for all values of the density,
the binding energy increasing with density to reach more than
half a GeV. The sharp peak, which occurs at chiral symmetry restoration, is
a consequence of the choice of a large scalar coupling, close to 
the value $g_s/g_\pi = 1$, for which the 
scalar diquark and the pion are degenerate. 
\section{The relativistic Faddeev equation}
Because of the separability of the NJL interaction, the 3-body relativistic
Faddeev equation in the ladder approximation can be reduced to an effective 
2-body Bethe-Salpeter 
equation describing the interaction between a quark and a diquark. Details
concerning the derivation of this equation can be found in the previous 
studies performed at zero density \cite{IBY95,HT94}. Explicitly, it is written:
\begin{equation}
\Psi(P,q) = \frac{i}{4\pi^4} \int d^4q' R(2P/3+q')V(q,q';P)\Psi(P,q') \quad ,
\label{Fad}
\end{equation}
where $R(2P/3+q')$ is the two-body $T$-matrix for the scalar diquark 
and $V(q,q';P)$ 
involves the product of the propagators of the spectator and exchanged quarks:
\begin{equation}
V(q,q';P)=\frac{(\gamma p_2' +m)(\gamma p_1' +m)}{({p_1'}^2-m^2)({p_2'}^2-m^2)}
 \quad .
\end{equation}
Here $p'_i$ (i=1,2,3) are the momenta of each valence quark, $P=p_1+p_2+p_3$ 
is the total momentum of the nucleon, and  $q,\, q'$ are the Jacobi variables 
defined by $p_3 \equiv  P/3-q$; $p_1' \equiv  P/3-q'$.  
We now look for the nucleon solution of (\ref{Fad}):
\begin{equation}
\Psi = \left( \begin{array}{c} \Phi_1(q_0,q) \\ \vec{\sigma}.\vec{q} \;
\Phi_2(q_0,q) \end{array} \right).
\end{equation}
 With this form for $\Psi$, Eq. (\ref{Fad}) becomes a set of two coupled 
integral equations. Following Huang and Tjon \cite{HT94}, we then perform a 
Wick rotation on 
the $q_0$ and $q_0'$ variables. This leads to two coupled 
complex integral equations which are 
solved iteratively \cite{HT94,MT69}. The initial guess for each of the wave 
functions
$\Phi_1(q_0,q)$ and $\Phi_2(q_0,q)$ consists of a Gaussian for the real
part and a derivative of a Gaussian for the imaginary part. The number of
iterations needed to reach convergence of the solutions is about four. 

The above discussion remains valid at finite density. Again, we incorporate the
effects of density by restricting the 3-momentum of each valence quark to 
values larger than the Fermi momentum $k_F$, i.e.:
\begin{equation}
 k_F \leq |\vec{p}_i\,'| \leq \Lambda \quad (i=1,2,3).
\end{equation}
This condition translates into Eq. (\ref{Fad}) as a complicated
cut-off on the 3-momen\-tum integration variable. 
 Apart from this restriction, 
the method of solving the Faddeev equation is the same as at zero density.     

\section{Results and conclusions}
At zero density we found that the nucleon is bound only if the scalar
coupling is strong enough, i.e. $g_s/g_\pi \geq 0.8$, which means
that the nucleon is not bound in either the
``standard'' NJL model ($g_s/g_\pi = 2/13$) or the 
colour-current interaction version ($g_s/g_\pi = 1/2$).  

\begin{figure}[ht]
\centerline{\psfig{figure=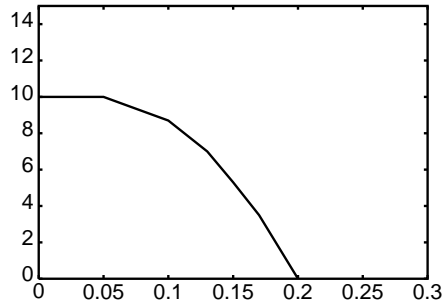,height=4.6cm}}
\caption{Nucleon binding energy (in MeV) as a function of $k_F/\Lambda$}
\label{bar}
\end{figure}

At finite density, we solve the Faddeev 
equation for the energy $E_{nuc}$ of the nucleon as a function of the Fermi 
momentum.
Results are shown in Fig.~\ref{bar} for
$g_s/g_\pi=0.83$. The binding energy of the nucleon, 
$B_{nuc}=E_{diq}+\sqrt{k_F^2+M^2}- E_{nuc}$, is depicted, again
as a function of the dimensionless variable $k_F/\Lambda$. The binding energy 
of the nucleon is
relative to the quark-diquark threshold: as there is no confinement in the 
NJL model, nothing can prevent the existence of a free diquark. In contrast to 
the diquark (shown in Fig.~\ref{diq} for the same value of $g_s/g_\pi$), 
which is
bound over the whole range of densities, the binding energy of the nucleon
decreases quickly with density and the binding disappears well before 
nuclear matter density (which corresponds to $k_F/\Lambda \simeq 0.45$). 


These results imply that, in the region 
characteristic of colour 
superconductivity (beyond chiral symmetry restoration), the quark Fermi sea
is unstable only with respect to formation of diquarks and not of 3-quark 
clusters. However, we have to emphasize 
that we included only the scalar part of the $qq$ interaction; at zero density,
several authors \cite{IBY95,Oet98} have shown that the axial-vector $qq$ 
interaction
gives an important contribution to the nucleon binding energy (of the
order 100 MeV), while the
axial-vector diquark is not bound for reasonable values of the axial-vector
coupling. Incorporating the axial-vector $qq$ interaction is necessary to 
obtain a more realistic picture of the evolution of the nucleon binding 
energy with 
density and could significantly modify the present results.

\end{document}